\documentclass[seceq,epsf]{ptptex}
\usepackage{wrapft}

\usepackage{graphicx}

\def\gtsim{\mathrel{\hbox{\raise0.2ex
\hbox{$>$}\kern-0.75em\raise-0.9ex\hbox{$\sim$}}}}
\def\ltsim{\mathrel{\hbox{\raise0.2ex
\hbox{$<$}\kern-0.75em\raise-0.9ex\hbox{$\sim$}}}}
\def\leqqq{\mathrel{\hbox{\raise0.2ex
\hbox{$<$}\kern-0.75em\raise-0.9ex\hbox{$=$}}}}
\def\geqqq{\mathrel{\hbox{\raise0.2ex
\hbox{$>$}\kern-0.75em\raise-0.9ex\hbox{$=$}}}}
\def \Romannumeral(#1) {\uppercase\expandafter{\romannumeral#1}}

\def \Romannumeral(#1) {\uppercase\expandafter{\romannumeral#1}}
\def \Rn(#1) {\uppercase\expandafter{\romannumeral#1}}
\def\Fig(#1){$${\overline{\underline{\rm Fig.{\ \ #1}}}}$$} 
\def\Tab(#1){$${\overline{\underline{
 \rm Table\ \  \uppercase\expandafter{\romannumeral#1}}}}$$} 

\def \D{\Delta}
\def \d{\delta}

\def \s{\sigma}

\def \o{ \omega }

\def \Kb{\overline K}

\def \fn1{N_{f_1}}
\def \fn2 {N_{f_2}}

\def \Romannumeral(#1) {\uppercase\expandafter{\romannumeral#1}}
\def \Rn(#1) {\uppercase\expandafter{\romannumeral#1}}

\def \qb{\overline q}
\def \sb{\overline s}
\def \Jb{\bar J}
\def \Kb{\overline K}

\def \pb{\overline p}

\def \db{\overline d}
\def \sb{\overline s}
\def \Bb{\overline B}
\def \Nb{\overline N}
\def \Hb{\overline H}
\def \Xb{\overline X}
\def \cb{\overline c}
\def \la{\langle}
\def \ra{\rangle}
\def \Nc{{\cal N}}
\def \Ncb{{\overline{ \cal N}}}
\def \10b{\overline {10}}
\def \3b{\overline {3}}

\def \mib #1 {$\mbox{\boldmath $ #1 $}$}
\def\sqr#1#2{{\vcenter{\vbox{\hrule height.#2pt \hbox{\vrule width.#2pt 
height#1pt \kern#1pt \vrule width.#2pt}\hrule height.#2pt}}}}

\title{
String Junction Model, Cluster Hypothesis,  Penta-Quark Baryon and Tetra-Quark Meson%
}

\author{
Masahiro  \textsc{IMACHI}$^{1}$
\footnote{E-mail:masaimac@tkg.bbiq.jp} ,
Shoichiro  \textsc{OTSUKI}$^{2}$
\footnote{E-mail:otk-s1ro@topaz.ocn.ne.jp} 
and
Fumihiko \textsc{TOYODA}$^{3}$
\footnote{E-mail:ftoyoda@fuk.kindai.ac.jp} 
\\ 
}

\inst{
$^1$ Yamagata University, ret.\\
$^{2}$ Kyushu University, ret.\\
$^{3}$Department of Electrical and Communication Engineering, \\
Kinki University of School of Humanity-Oriented Science and Engineering
}

\abst{
Thirty years ago we proposed string junction model of hadrons and examined structure and reaction of hadrons including exotic ones. Mass $m$ of exotic  hadrons of light  quarks is roughly given by $m \sim N_J\cdot m_B$,  where $N_J$ is the total number of junctions and $m_B \sim 1$ GeV is the ordinary light baryon mass.  In this paper we introduce ``cluster hypothesis"  into the model  by which   mass of a complex hadron is given by the sum of masses of clusters composing it. The hypothesis guarantees the established picture that  mass differences of hadrons of the same string junction structure are due to those of  the constituent quarks. \par
A candidate for penta-quark baryon $\Theta$(1530 MeV, $S=+1)$  including a strange anti-quark ${\sb}$ and that for tetra-quark meson $Z^+$(4430 MeV)  recently  reported by the Bell collaboration are examined in parallel. $\Theta $ is considered to have non-strange partners, which are lighter by the mass difference $\Delta_s$ between strange and non-strange quarks. Mass of such light penta-quark baryons with $N_J=3$ is expected to be about 3 GeV.  Several parameters of the model are estimated such as mass of junction of $m_J \sim O(10)$ MeV. While mass of light tetra-quark meson with $N_J=2$ is expected to be about 2 GeV, $Z^+$(4430 MeV) containing $(u,c,{\db},{\cb})$ gives a  clue to determine some parameters of the model, e.g., inter-junction string energy $m_{IJ}$.  
 }

\begin{document}

\maketitle

\section { \bf Introduction} 
There exist numbers of ordinary hadrons made from quark $q$ and anti-quark $\qb$: the ordinary meson $M=q \qb$ and the ordinary baryon $B=qqq$. A discovery of unconventional hadrons with extraordinary structure, however,  had not been reported for a long time. This poses a quite keen contrast with  situations of atoms and atomic nuclei both as physical composite systems. The reason of the contrast  may be closely related to the mechanism of quark confinement. Since we have almost no established methods to treat low energy phenomena of QCD, we adopt in this paper 
a model based on the string picture of confinement.  We expect this will give a clue to clarify low energy nature of QCD in a complementary way to other approaches  such as  lattice gauge theory. \par 
In our model  
 quark confinement is considered to be realized by colored string. The string has ``orientation"  because, when the string is cut by pair creation of quarks, the sequence of $q$ and $\qb$ is unique. When we define the orientation by the direction toward a confined quark at the end, there should exist in the baryon a ``singular point" from which  three strings emerge and where  three colors are neutralized. This point is called `` junction" \cite{IOT1}. In  the string junction model we proposed in 1977\cite{IOT} (abbreviated as SJM)(see also \citen{sakita},\citen{A},\citen{RV}), we  investigated unconventional hadrons including exotic ones. The reason why they were so difficult to be observed was attributed to their complex structure, in particular, to the nature of junction, which was regarded as a physical entity similarly to quarks.\par
In 2003 a discovery of an exotic penta-quark baryon,  $\Theta(1530)$ with strangeness $S=+1$,   was reported\cite{Nakano}, though  experimental judge for this candidate is  controversial up to now. 
Recently the Belle Collaboration has reported a new resonance $Z^+(4430)$\cite{Belle}. As  emphasized in the paper, $Z^+$  is a charged particle and is a good candidate for tetra-quark meson containing $(u, c, \db, \cb)$. In the present article, we investigate  exotic hadrons of SJM. 

Our paper is organized as follows. 
In section 2, a  brief summary of hadrons of SJM is firstly given. 
Except for the ordinary meson, baryon and a simple string ring,  every unconventional hadron has a chain of junctions and anti-junctions as its skeleton. Under reasonable assumptions the standard mass $m$ of a hadron containing only light quarks is roughly given by
\begin{eqnarray}
%
m \sim N_J \cdot m_B,
\label{} 
\end{eqnarray}%
\noindent where $N_J$ is the total number of junctions of the hadron and $m_B \sim$ 1 GeV is the ordinary light baryon mass. Then the mass of light penta-quark baryon with $N_J=3$ is supposed to be about 3 GeV while that of light tetra-quark meson with $N_J=2$ to be about 2 GeV.\par
We introduce ``cluster hypothesis"  into SJM. Namely, 
when all the inter-junction strings within a hadron are cut at the same time, each connected part  is called as a single cluster. By the hypothesis mass of a hadron with multi-cluster structure is given by the sum of masses of the clusters,
and the traditional picture that  mass differences of hadrons with the same string junction structure come from those among constituent quarks is guaranteed.  For ordinary mesons and baryons of light quarks, the mass difference between strange and non-strange quarks is expressed by the mass formula of Gell-Mann-Okubo(GMO)\cite{G1}\cite{O1} with $T^3_3$ breaking of flavor $SU(3)_f$.  We consider  $T^3_3$ breaking should be applied to each cluster but not to any unconventional hadrons with complex structure as a whole.  A comparison of  exotic hadrons of SJM with those of  Skyrmion model\cite{Diak}\cite{Prasz} and di-quark model\cite{JW} is also given. \par
In section 3 we discuss quark rearrangement diagram, duality and selection rules, where physical roles of junction are shown in connection with tetra-quark meson and penta-quark baryon. Although   ``indirect",  $\Bb$-$B$ duality gives important information that the mass of light tetra-quark meson should  be about $2m_B$. 
 The selection rule by Freund-Waltz-Rosner\cite{FWR} to suppress  the decay of tetra-quark meson into two ordinary mesons is replaced by a  rule to forbid junction  hair pin line diagram\cite{IO2}, which is in parallel to the OZI rule\cite{O2}\cite{Z}\cite{Iiz} to forbid quark  hair pin line diagram. This would explain the ``narrow" width of $Z^+$.  \par
In section 4, the mass of junction is estimated as $m_J \sim O(10)$ MeV. A heuristic argument by uncertainty relation is  given on the mass of string attached to a source quark together with its length in the ground state hadron. 
When $Z^+(4430)$ is established, it helps  to determine basic parameters of SJM  in a  ``direct" way.
In subsection  4.4 newly added after the report of $Z^+$, we identify it with 
tetra-quark meson composed from ($u,c,J;{\Jb} ,{\db},{\cb}$) including a pair of junctions, and  estimate the mass $m_{IJ}$ of inter-junction string  and the energy $ \delta $ necessary to cut it by creating a pair of quarks. 
This  makes it possible to predict   
masses of  unconventional hadrons, such as  3380 MeV for $\Theta$ containing  an $\sb$. A summary of the parameters of SJM is presented in section 5. 

\section { \bf String junction model, cluster hypothesis and mass of hadrons } 
\subsection{\bf String junction model}
\begin{figure}
 \epsfysize= 9 cm
\hskip.05cm
\epsfbox{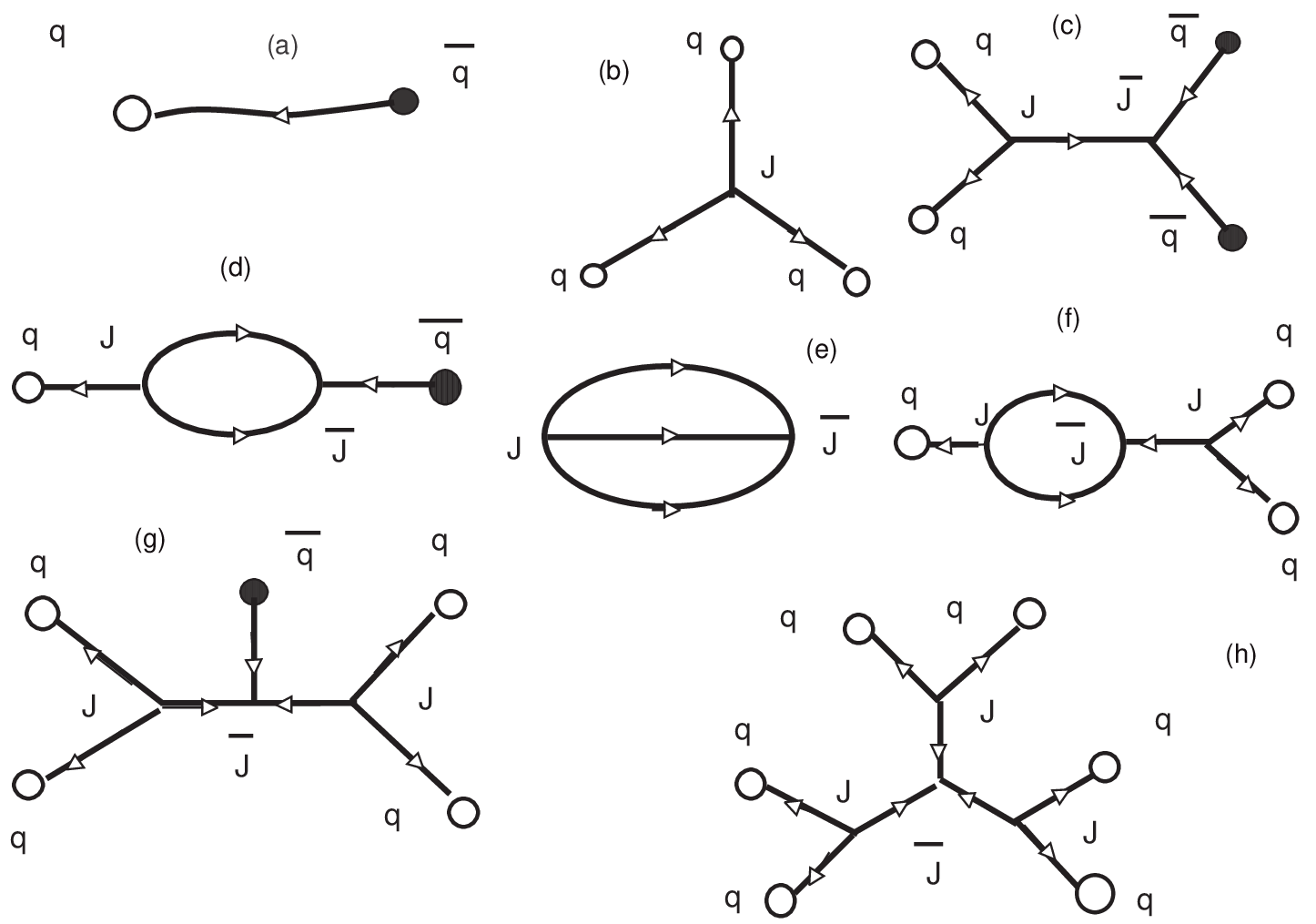}

   \caption{Hadrons in string junction model. (a)ordinary meson $M^0_2$, (b)ordinary baryon $B^1_3$, (c)exotic meson 
   $M^2_4$, (d)$M^2_2$, (e)gluonium $S^2_0$, (f)$B^3_3$, (g)exotic baryon $B^3_5$,
    (h)di-baryon $D^4_6$. Upper suffix and lower suffix denote the total numbers of junctions and quarks respectively. }
   \label{fig1}
   \end{figure}

In SJM there are various hadrons such as shown in Fig. {\ref{fig1}}. In this paper, the upper suffix of hadron symbols denotes $N_J$, the total number of junctions (sum of junction number and anti-junction number), and the lower suffix denotes $N_q$, the total number of quarks (sum of quark number  and anti-quark number).

\input epsf.tex
%
\begin{itemize}
\item[1.]Meson octet $M(8)$   and nonet $M(9)=M(8) \oplus  M(1)$ have the structure $M\equiv M^0_2$ in which $q$ and  $\qb$ are connected with each other by an oriented string(Fig. 1(a)). 
Baryon octet $B(8)$ and decuplet $B(10)$ have the structure $B\equiv B^1_3$, in which three quarks are connected with a junction $J$(Fig.1(b)).\par

\item[2.]A meson whose quantum number is realized only by four quark system $qq\qb \qb$ is expressed as $M^2_4$ in SJM. This exotic meson contains a pair of junctions ($J$  and $\Jb$), and $qq$ and $\qb \qb$ are connected with $J$ and $\Jb$ respectively. Junction and anti-junction are connected by an ``inter-junction string"(Fig.1(c)) .   Exotic baryon $B^3_5$ contains $qq\qb qq$ and $J \Jb J$, and the latter are connected by two inter-junction strings. Quarks $qq$ and anti-quark $\qb$ are connected with $J$ and   $\Jb$ respectively(Fig. 1(g)).\par

\item[3.] A  string ring which contains neither quark nor junction is called ``Pomeron". The exchange of this in $t$-channel leads to diffraction scattering at high energies.\par

\item[4.]Except for  ordinary mesons,  baryons and  Pomeron,  hadrons   in SJM contain at least one inter-junction string.  
It should be noted that unconventional hadrons have such skeleton type structure  of multi-junction.  $M^2_2$ (Fig. 1(d)), $S^2_0$ (Fig.1(e)) and $B^3_3$ (Fig.1(f)) are examples of unconventional hadrons.\par
\item[5.]In SJM, di-baryon $D^4_6$ is not merely six quark state but it has the structure with three junctions and one anti-junction(Fig1(h)).\par
\end{itemize}

\vskip 0.5cm
\subsection {\bf  Cluster hypothesis and mass of hadrons } 
Now we will introduce  ``cluster hypothesis" into SJM as  (I) to  (IV) below: 
\begin {itemize}
\item[(I)]The ordinary meson $M$ and  baryon $B$, which do not contain inter-junction string, are defined   as a single cluster. \par
\item[(II)]{\bf When all the inter-junction strings within a hadron are cut at the same time, each connected part is defined as ``a single cluster". Let $N_{IJ}$ be the number of inter-junction string connecting $J$ and $\bar J$. Any hadron with $N_{IJ} \neq 0$ has multi-cluster structure.} \par
\item[(III)]It may be natural in SJM to express mass of hadron as\cite{IO1}\cite{Im_kik}\cite{Im_but}\par
\begin{eqnarray}
m=m_q N_q+m_J N_J+m_{IJ} N_{IJ},
\label{masshad} 
\end{eqnarray}%
where $m_q$ denotes the ``constituent quark mass", which includes both the mass $\tilde m_q$ of the ``source quark" and the energy of the string connecting it with a junction. (A relation between $m_q$ and  $\tilde m_q$ is discussed in subsection 4.2.) When we assign $N_{IJ}=1/2$ to each of the ``cut" inter-junction string, mass of any multi-cluster hadron is given by the sum of mass of the clusters (``additivity"):
\begin{eqnarray}
m\sim \sum_{\rm cluster} m_{\rm cluster}.
\label{additivity} 
\end{eqnarray}%
\item[(IV)] Since the mass of junction and the mass of inter-junction string are flavor independent, the mass of multi-cluster hadron  faithfully reflects the constituent quark mass. \par
\end{itemize}

\begin{table}
\caption[]{Mass of  clusters}
\label{masscluster}
\begin{center}
\begin{tabular}{| c |c  |c |c | c |} \hline 
cluster                 & $N_q$ &$N_J $ &$N_{IJ} $ &  $m_{\rm cluster}$  \\ \hline
$M^0_2(\qb q)$           & 2& 0 & 0               & $2m_q$                      \\
$B^1_3(qqJq)$            & 3& 1 & 0               & $3m_q+m_J$                  \\ \hline
                         &  &   &                 &                             \\
 $qqJ \cdot $(HIJ)     & 2&  1&1/2              &$2m_q+ m_J+1/2 \cdot m_{IJ}$ \\ \hline
                         &  &   &                 &                             \\
$qJ \cdot$ 2(HIJ)    & 1& 1 & $1/2\times 2$   & $m_q+ m_J+  m_{IJ}$         \\ \hline
                         &  &   &                 &                             \\
$J \cdot$ 3(HIJ)    & 0&1  & $1/2\times 3 $  &   $m_J+3/2 \cdot m_{IJ} $   \\ \hline
\end {tabular}
\end{center}
\end{table}
\vskip 0.5cm


Except for the ordinary meson, baryon and Pomeron, there are three kinds of clusters: $\{ qqJ \cdot ({\rm {HIJ}} ) \}$, $\{ qJ \cdot 2({\rm {HIJ} }) \}$ and  $\{ J \cdot 3({\rm {HIJ} }) \}$ together with their charge conjugate partners, where HIJ is the abbreviation of one half of the inter-junction string. Mass of them is shown in Table \ref{masscluster}. Several examples of cluster structure including exotic hadrons are given in the first column of Table \ref{uncvhadron}.  For examples, exotic meson $M^2_4$ is composed of two clusters, \{$qqJ \cdot $(HIJ)\} and   $\{\qb \qb \Jb\cdot  $(HIJ)\}, and 
 $S^2_0$ represents the gluonium without quark and is composed of two clusters \{$J \cdot $3(HIJ)\}, \{$\Jb \cdot $3(HIJ)\}.
 Penta-quark baryon $B^3_5$ is composed of three clusters,   \{$qqJ \cdot $(HIJ)\},   \{$\qb \Jb \cdot $2(HIJ)\} and  \{$qqJ \cdot $(HIJ)\}. Likewise, $B^3_3$ is composed of three clusters, \{$qqJ \cdot $(HIJ)\}, \{$\Jb \cdot $3(HIJ)\}  and \{$q J \cdot $2(HIJ)\}.

The mass of hadrons of Eq. (\ref{masshad}) is rewritten as

\begin{eqnarray}
m = m_B  \cdot N_J-\delta \cdot N_{IJ}
\label{} 
\end{eqnarray}%
if we note the relation
\begin{eqnarray}
 N_q=3 N_J -2 N_{IJ},
\label{} 
\end{eqnarray}%
 where 
\begin{eqnarray}
 m_B=3 m_q+m_J
\label{} 
\end{eqnarray}%
denotes the mass of the ordinary baryon $B$  and 

\begin{eqnarray}
 \delta=2 m_q-m_{IJ}
\label{} 
\end{eqnarray}%
means  that the energy necessary to cut an inter-junction string by quark pair creation is  $\d$.\par

\begin{table}
\caption[]{Standard mass of unconventional hadrons}
\label{uncvhadron}
\begin{center}
\begin{tabular}{| l |c | c |c |c |c |c| } \hline 
Unconv. Meson            & $B$   &  $N_q$  &$ N_J $&$ N_{IJ} $ & mass            &cluster structure \\ \hline
                         &       &         &       &           &                 &\\ 
$M^2_4(qqJ;{\Jb}\qb \qb)$ &  0    &    4    &    2  & 1         & $2 m_B-\delta$  &2 clusters\\ \hline
                         &       &         &       &           &                 &\\ 
$M^2_2(qJ ;\Jb \qb )$     &  0    &   2     &    2 &  2         & $2 m_B-2\delta$ &2 clusters\\ \hline
                         &       &         &       &           &                 &\\ 
$S^2_0(J ;\Jb )  $        &  0    &    0    &  2   &  3         &  $2 m_B-3\delta$&2 clusters\\ \hline\hline
Unconv. Baryon     & $B$ &  $N_q$&  $ N_J $ &  $ N_{IJ} $ & mass &\\ \hline
                         &       &         &       &           &                 &\\
$B^3_5(qqJ;\qb{\Jb}; J qq)$& 1     &  5      &    3  & 2         & $3 m_B-2\delta$ &3 clusters\\ \hline
                                                                                   
                         &       &         &       &           &                 &\\
$B^3_3(qqJ ;{\Jb} ;J q )$  & 1     & 3       &    3  &  3        & $3 m_B-3\delta$ &3 clusters\\ \hline\hline
Di-baryon        & $B$ &  $N_q$& $ N_J $&$ N_{IJ} $ & mass            &\\ \hline
                         &       &         &       &           &                & \\
$D^4_6(qqJ;{\Jb};Jqq;Jqq)$  & 2     &  6      &    4  & 3         & $4 m_B-3\delta$&4 clusters \\ \hline
\end {tabular}
\end{center}
\end{table}

The cluster structure  of unconventional hadrons in Fig. \ref{fig1} and standard mass of them are shown in  Table \ref{uncvhadron}.  \par
It may be natural to assume parameters $m_q, m_J, \d, m_{IJ} $ are ``universal" in SJM irrespective of hadron structure. If the parameter $\d$ is small compared with $m_B$, multi-cluster hadron mass is approximately given by
\begin{eqnarray}
 m\sim N_J\cdot m_B.
\label{} 
\end{eqnarray}%
Actually, we will see in  section 3 and subsections 4.3 and 4.4 that  the parameter $\d$ is rather small. Then mass of tetra-quark meson and mass of penta-quark baryon are $\sim 2 m_B$ and $\sim 3 m_B$, respectively. \par

\subsection{\bf Mass splitting due to strange quark}
Mass splitting due to strangeness among hadrons belonging to an irreducible $SU(3)_f$ representation is given by $T_3^3$ breaking of the GMO mass formula as\par
\begin{eqnarray}
 m=a + b Y+c\{ I(I+1)-\frac{1}{4}Y^2 \},
\label{GMO} 
\end{eqnarray}%
where $I$ denotes isospin and $Y \equiv B+S$ does hypercharge, $B$ and $S$ being baryon number and strangeness respectively. 
The physical content of the $T^3_3$ breaking is clearly understood by the famous equal mass splitting of decuplet baryon $B(10)$. For decuplet representation $I=Y/2+1$, so Eq.(\ref{GMO}) is a linear function of $Y$ or $S=Y-1$ and mass splitting of hadron is reduced to that of the constituent quarks:\par
\begin{eqnarray}
%
\Delta_s=m_s-m_u \sim m_s-m_d.
\label{Delta} 
\end{eqnarray}%
 \par

For the cluster $\{ qqJ \cdot ({\rm {HIJ}} ) \}$, there are two cases for the quark pair $qq$. For the anti-symmetric case $[q,q]$ which belongs to ${\bf {\3b}}$ representation of $SU(3)_f$, $I=-Y/2+1/3$ while for the symmetric case$\{ q,q\}$ which belongs to ${\bf 6}$ representation, $I=+Y/2+2/3$. Again Eq.(\ref{GMO}) is linear on $S$ for either case. For completeness, note that the cluster \{$q J \cdot $2(HIJ)\} of course belongs to $\bf 3$ representation, thus has linear dependence on $S$. In this way, the mass splitting of  hadrons in SJM under the cluster hypothesis is always reduced to the number of $s$ and $\sb$  quarks contained.\par

\vskip 0.5cm

The configuration of penta-quark baryon $B^5_3$ is
\begin{eqnarray}
%
|(qq)J;\Jb \qb; (qq)J\ra
\label{} 
\end{eqnarray}%
where $(q,q)$ denotes either $[q,q]$ or $\{ q,q\}$. For simplicity we take the same mass of $2m_q$ for the both cases. If we write concretely, we have 
\begin{eqnarray}
\left\{
\begin{array}{lll}
|(\Nc \Nc) J;\Ncb \Jb;(\Nc \Nc)  J\ra  &(S=0), & m=3m_B-2\d \\ 
|(s \Nc) J; \Ncb \Jb; (\Nc \Nc) J\ra   &(S=-1), & m=3m_B-2\d +\D_s\\ 
|(ss) J; \Ncb \Jb; (\Nc \Nc)  J\ra   &(S=-2),  &m=3m_B-2\d +2\D_s\\ 
|(s\Nc) J; \Ncb \Jb; (s \Nc)  J\ra   &(S=-2),  &m=3m_B-2\d +2\D_s\\ 
|(ss)J; \Ncb \Jb;(s \Nc)  J\ra  &(S=-3),  &m=3m_B-2\d +3\D_s\\ 
|(ss)J; \Ncb \Jb;(ss) J\ra  &(S=-4),  &m=3m_B-2\d +4\D_s\\ 
|(\Nc \Nc) J;\sb \Jb; (\Nc \Nc) J\ra &(S=+1),&  m=3m_B-2\d +\D_s\\ 
|(s \Nc) J; \sb \Jb;(\Nc \Nc ) J\ra &(S=0), & m=3m_B-2\d +2\D_s\\ 
|(ss) J;\sb \Jb;(\Nc \Nc)  J\ra  &(S=-1), & m=3m_B-2\d +3\D_s\\ 
|(s \Nc) J;\sb \Jb;(s \Nc)  J\ra  &(S=-1), & m=3m_B-2\d +3\D_s\\ 
|(ss)J;\sb \Jb ;(s \Nc)  J\ra &(S=-2), & m=3m_B-2\d +4\D_s\\ 
|(ss) J;\sb \Jb; (ss) J\ra  &(S=-3), & m=3m_B-2\d +5\D_s. 
\end{array}
\right.
\label{B5} 
\end{eqnarray}%
For the pair of same flavor quarks, $uu, dd$ and $ss$, only the  symmetric combination $\{q,q \}$ survives.\par
Characteristic features of penta-quark baryon of SJM are as follows:\par
\begin{itemize}
\item[(1)]If we neglect  both $\Delta_s$ and $\delta$, which may be  expected to be small compared with $m_B$, exotic baryon $B^3_5$ will have mass 
around three times of baryon mass: 
\begin{eqnarray}
 m_{B^3_5} \sim 3m_B \sim 3{\rm   \ GeV}.
\label{penta quark baryon} 
\end{eqnarray}%
\item[(2)]If we take into account the mass breaking $\Delta_s= m_s-m_u \sim m_s-m_d
 = $ 130 $\sim$150   \ MeV\footnote{For example, 
$\Delta_s \sim (2m_{u, d}+m_s)-(3m_{u, d}) \sim (2m_{\Lambda}+6m_{\Sigma}+12m_{\Sigma}^*)/{20}-(4m_N+16m_\Delta)/{20}=1300-1170=130$   \ MeV.}, we have
\begin{eqnarray}
%
m({|(\Nc \Nc) J; \Ncb \Jb; ( \Nc \Nc) J\ra})<m({\Theta=|(\Nc \Nc) J;  \sb \Jb; (\Nc \Nc) J\ra}), 
\label{} 
\end{eqnarray}%
with the mass difference  $ \D_s$. \par
\item[(3)]In SJM, $\Theta$ can be neither the lightest of penta-quark baryons  nor be so light as 1530   \ MeV.\par
\end {itemize}
\vskip 0.5cm
Here we will add short comments on some other models of penta-quark baryon;  Skyrmion model and di-quark model of Jaffe and Wilczek. \par
In Skyrmion model\cite{Diak}\cite{Prasz},   penta-quark baryon is regarded as a single topological soliton. The soliton including  $\Theta$ is assigned to pure ${\bf {\10b} }$ representation of $SU(3)_f$.  Mass of  $\Theta$  is calculated to be 1530 MeV. The quark contents of the  ${\bf {\10b} }$ representation together with the average of the total number of $s$ quark for this representation are given below;
\begin{eqnarray}
\left\{
\begin{array}{cll}
  \Theta &\sim   |[ud][ud]\sb \ra \ra , & \la N_s\ra=1\\ 
  N &\sim\{|[ud][ud]{\cal \Nb} \ra+|[ud][ {\cal N}s]\sb \ra+ |[{\cal N}s][ud]\sb\ra \}/\sqrt{3},& \la N_s\ra=4/3\\ 
  \Sigma &\sim \{|[ud][{\cal N}s]{\cal \Nb} \ra+|[{\cal N}s][ud]{\cal \Nb} \ra+|[{\cal N}s][{\cal N}s]\sb \ra\}/\sqrt{3}, & \la N_s\ra=5/3\\ 
  \Xi &\sim |[{\cal N}s][{\cal N}s]{\cal \Nb}\ra,  &\la N_s\ra=2.
\end{array}
\right .
\label{10bar} 
\end{eqnarray}%
Among them, $\Theta$ is the lightest,  but the mass differences are only $(1/3)\Delta_s$. In the di-quark model\cite{JW},  a pair of quarks  is a constituent unit of hadrons, 
so that the configuration of penta-quark baryon is (di-quark)-(di-quark)-(anti-quark). The resulting ${\bf \10b}$ and ${\bf 8}$ representation are considered to be ideally mixed. The lightest state in this scheme is $|[ud][ud]\Nb\ra$, which is identified with Roper resonance(1440 MeV). The state $\Theta=|[ud][ud]{\sb}\ra$ is heavier than  this by $\Delta_s$. Masses  of $\Theta$ of  the two models  are about 1.5 GeV. \par



\section { \bf Quark rearrangement diagram, duality and selection rules -----physical roles of junction} 
That hadron structure and hadron reaction are strongly related is clearly seen through duality and is well manifested by quark rearrangement diagrams\cite{IMNS} \cite{J}. Except for  diffraction scattering corresponding to Pomeron exchange, presence or absence of resonances in $s$-channel and Regge poles exchanged in $t$-channel are dual.\par\noindent
\subsection{\bf Quark rearrangement diagram and duality}\par
\noindent 1){\bf ${\Kb} N$  and $KN$  duality and penta-quark baryon}\par
\begin{figure}
 \epsfysize= 4 cm
\hskip.05cm
\epsfbox{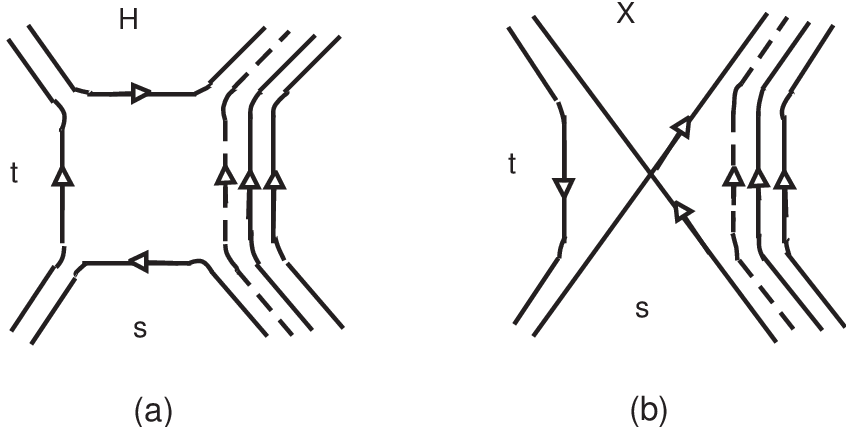}

   \caption{(a)$\Kb N$ scattering diagram. Quarks are drawn by full lines. Junction is drawn by a dashed line. It is the $H$ type diagram and is $s$-$t$-channel dual. (b)   $ K N$ scattering diagram. It is the $X$ type diagram and is $u$-$t$-channel dual. So it does not contain $s$-channel baryon resonance. }
   \label{fig2}
   \end{figure}

 The non-Pomeron part of  $\Kb N$ scattering  is given only by $H$ type diagram,  while the non-Pomeron part of $KN$ scattering is only by $X$ type one. Quark rearrangement diagram of  $\Kb N$ and  $K N$  is depicted in Fig. \ref{fig2}.  Intermediate state of $s$-channel of the former is contributed from ordinary baryons with $S=-1$,  while that of the latter with $S=+1$ is never from penta-quark baryons but simply the ones obtained by $s$-$u$ crossing of Fig.\ref{fig2}(a).  Note that penta-quark baryon has the structure not only of five quarks but it has three junctions($J\Jb J$) in SJM and never appears in $s$-channel of $KN$.\par
\input epsf.tex
%

Total cross section difference $\Delta \sigma_T(KN)$ is related to the imaginary part of forward scattering amplitude $T$ by optical theorem as below.\par
\begin{eqnarray}
%
\begin{array}{rl}
\D \s_T(KN)&\equiv \s_T({\Kb} N)- \s_T(KN)=(1/s)\left({\rm Im}T({\Kb}N)-{\rm Im}T(KN)\right )\\
&=(1/s)({\rm Im}T(H)-{\rm Im}T(X))\sim (1/s){\rm Im}T(H) .
\end{array}
\label{KNtotal} 
\end{eqnarray}%
\noindent Here Pomeron contribution of diffraction scattering is cancelled between  $\Kb N$ and $ K N$ scatterings. 
Thus  all the imaginary parts Im$T$ in Eq. (\ref{KNtotal}) should be understood to be the imaginary part of the respective ``non-Pomeron" amplitude.
That Im$T(KN)\equiv$ Im$T(X) \sim 0$ strongly indicates absence of resonances in $s$ channel on the one hand and corresponds to exchange degeneracy of meson Regge poles in $t$ channel on the other.  Then $\Delta\sigma_T (KN)$ is due to Im$T({\Kb}N)\equiv$ Im$T(H)$.  This imaginary part owes to numbers of $s$ channel resonances with $S=-1$ above ${\Kb}N$ threshold on the one hand, whose average coincides with $t$ channel Regge poles known as finite energy sum rule\cite{Igi}\cite{DHS}.  This  $\Delta\sigma_T (KN)$ phenomenologically decreases as $s^{-1/2}$, where $s$  is the square of the $s$ channel energy. The factors $(1/s)$ in Eq.(3.1) are kinematical flux factor, and $s$ dependence of  Im$T(H) \sim s^{1/2}$ is given by the intercept of the meson Regge trajectories which, in turn, is interpreted by the number of the exchanged constituents, $N_q = 2$.  Baryon resonances mediate only in $u$ channel of $KN$ scattering. \par
 
Idealized duality is elegantly expressed by Veneziano amplitude\cite{V}; $s$-$t$ dual Veneziano amplitude $V(s, t)$  corresponds to $H$ type diagram and  $u$-$t$  dual one $V(u, t)$ to $X$ type one in Fig. \ref{fig2}, as summarized in  Table \ref{duality}.\par

\begin{table}
\caption[]{Duality of meson-baryon scattering}
\label{duality}
\begin{center}
\begin{tabular}{| c | c |c |} \hline 
                    & $H$ type   &  $X$ type       \\ \hline
                    &            &                 \\
$s$-channel resonances   &  yes       &   no            \\ \hline
                    &            &                 \\
${\rm Im} \ T(t=0)$      &   yes      &   no            \\ \hline
                    &            &                 \\
${\rm Re} \ T(t=0)$    &   no       &   yes           \\ \hline
                    &            &                 \\
duality             &$s$-$t$ dual&   $u$-$t$ dual  \\ \hline
                    &            &                 \\
Veneziano amplitude & $V(s, t)$  &  $V(u, t)$     \\ \hline
\end {tabular}
\end{center}
\end{table}


\noindent 2) {\bf $\Nb N$  and $NN$  duality and unconventional mesons}\par
\begin{figure}
 \epsfysize= 12 cm
\hskip.05cm
\epsfbox{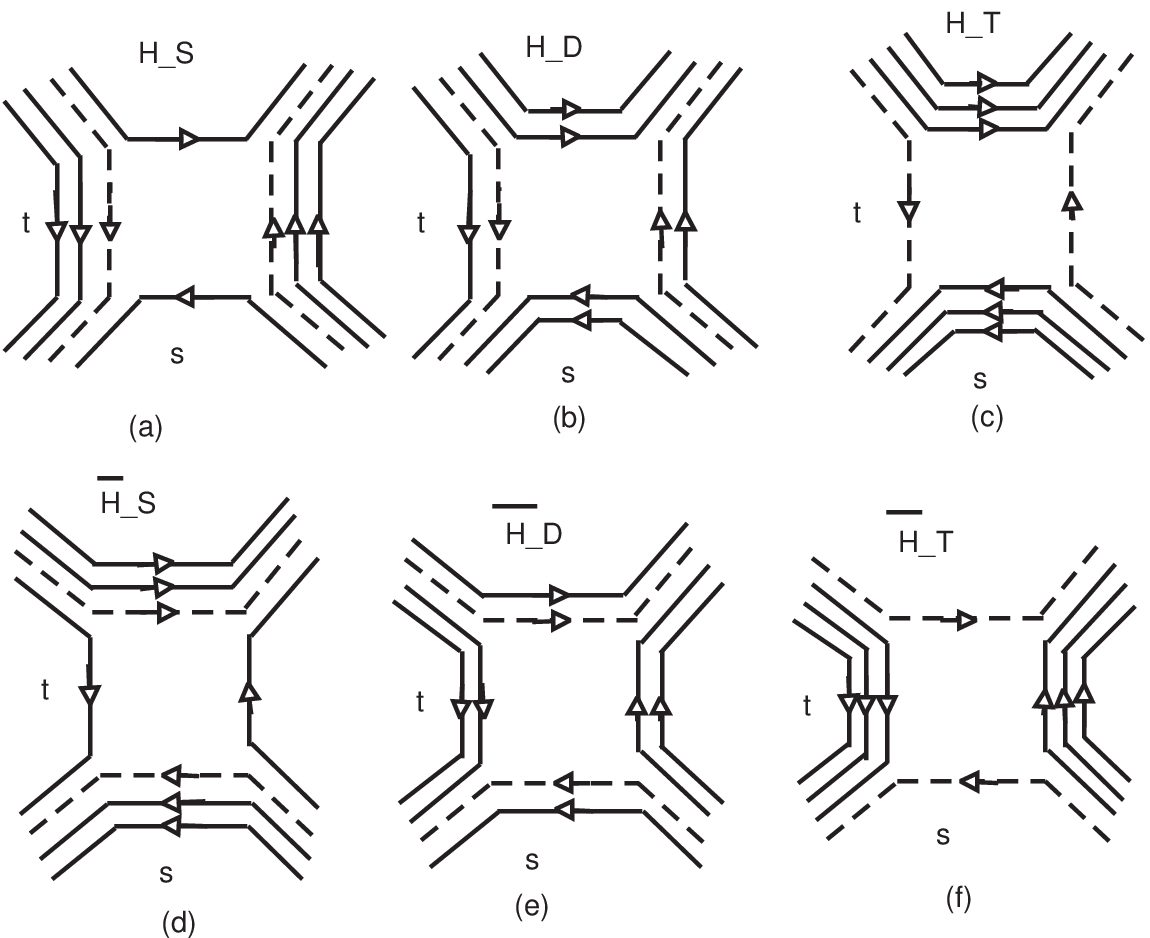}

   \caption{ $H$ type diagrams in ${\Nb}N$ scattering. Quarks are drawn by full lines. Junctions are drawn by dashed lines.  (a)$H_S(t:M, s:M^2_4)$, (b)$ H_D(t:MM, s:M^2_2)$,(c)$ H_T(t:MMM, s:S^2_0)$, and their $s$-$t$ crossed diagrams:(d)${\Hb}_S(s:M, t:M^2_4)$,  (e)${\Hb}_D(s:MM, t:M^2_2)$, (f)${\Hb}_T(s:MMM, t:S^2_0)$ }
   \label{fig3}
   \end{figure}

Corresponding to Fig. \ref{fig2}(a) of $\Kb N$ scattering, there are three quark rearrangement diagrams of $H$ type in SJM\cite{IOT} of $\Nb N$ scattering:\par\noindent
\begin{eqnarray}
%
 H_S(t:M, s:M^2_4),  H_D(t:MM, s:M^2_2), H_T(t:MMM, s:S^2_0).
\label{H_nn} 
\end{eqnarray}%
\noindent Note that the existence of junction leads to new diagrams $ H_D(t:MM, s:M^2_2)$ and  $H_T(t:MMM, s:S^2_0) $ in $\Nb N$ scattering. 
Here $S,D,T$ mean respectively single, double and triple exchange of ordinary mesons  in their $t$-channels. Note the contribution of unconventional mesons in their $s$-channels. 
There are also three    $s$-$t$ crossed ones;
\begin{eqnarray}
%
 {\Hb}_S(s:M, t:M^2_4),  {\Hb}_D(s:MM, t:M^2_2), {\Hb}_T(s:MMM, t:S^2_0).
\label{} 
\end{eqnarray}%
These six  $H$ type diagrams are shown in Fig. \ref{fig3} (a) $\sim$ (f).\par
We examine  difference of total cross section $\D \s_T (NN)$ expressed as below:
\begin{eqnarray}
%
\begin{array}{rl}

&\D \s_T(NN)\equiv \s_T({\pb} p)- \s_T(pp)=(1/s)\left\{{\rm Im}T({\pb}p)-{\rm Im}T(pp)\right \}\\
&=(1/s)\left\{{\rm Im}(T(H_S)+T(H_D)+T(H_T)+T({\Hb}_S)+T({\Hb}_D)
+T({\Hb}_T))\right.\\
&-\left.{\rm Im}(T(X_S)+T(X_D)+T(X_T)+T({\Xb}_S)+T({\Xb}_D)+T({\Xb}_T))\right \}\\
&\sim (1/s) {\rm Im}(T(H_S)+T({\Hb}_T)).
\end{array}
\label{delsigB} 
\end{eqnarray}%



\input epsf.tex
\noindent Similarly to Eq.(\ref{KNtotal}), Im$T$ denotes the respective non-Pomeron amplitudes.\par

A few comments are in order. They are essentially the same as those on $\D \s_T(KN)$.
Non-Pomeron part of $T(pp)$ is known to be almost real. It is given by six $X$-type diagrams, which are analogous to $X$ type diagram in Fig.\ref{fig2}(b) of $KN$ scattering. They are respectively obtained from six $H$ type diagrams in Fig. \ref{fig3} by $s$-$u$ crossing, whose $s$-channel states are never di-baryon resonances. Note that ``di-baryon" in SJM is composed not only from six quarks but also from three junctions and one anti-junction, as shown in Fig.\ref{fig1}(h). Di-baryon, thus,  does not contribute to $X$ type diagrams in $N N$ scattering.
 Two of six diagrams,  $H_S (t:M, s:M_4^2)$ and  ${\Hb}_T (s:MMM, t:S_0^2)$,   are supposed to give the leading  contribution with $s^{1/2}$ to   Im $T(\pb p)$, since the least number of constituents are exchanged in $t$-channel: $(q,\qb)$-exchange in  $H_S (t:M, s:M_4^2)$ and  $(J, \Jb)$-exchange\cite{IOT}\cite{EH}  in ${\Hb}_T (s:MMM, t:S_0^2)$. \par

 Although it is  indirect evidence, duality gives important information about the mass of tetra-quark mesons. 
That ${\Delta}\s_T(NN)$ is phenomenologically large ``above ${\pb}p$ threshold" would surely indicate contributions of tetra-quark mesons $M_4^2$ in the $s$ channel of $H_S(t:M,s:M_4^2)$ and imply that many of their masses are above but near to $2m_B$ or that $\delta$ is small negative or positive. But  continuum  states $MMM$ of ${\Hb}_T(s:MMM,t:S_0^2)$ also give the leading contribution, which would mask the possible separated  resonance peaks of $M_4^2$.


\par
\subsection{\bf Selection rules and  penta-quark baryon} \par
\begin{figure}
 \epsfysize= 4 cm
\hskip.05cm
\epsfbox{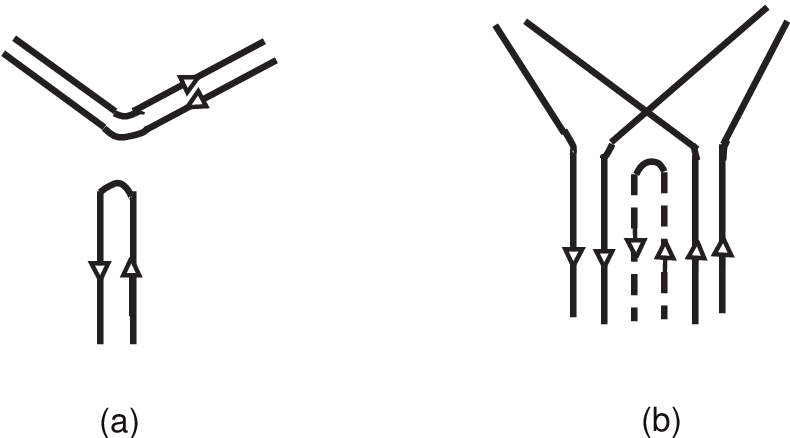}

   \caption{(a) is OZI forbidden decay(e.g., $\phi \to \rho \pi$).  It contains quark hair pin diagram. (b) is the process $M^2_4\to MM$ and is  ``junction hair pin (FWR) rule" forbidden decay. It contains junction hair pin diagram.  Junction is drawn by a dashed line.  $B^3_5\to MB$ occurs through a similar process. }
   \label{fig4}
   \end{figure}

 
In order to secure a suppression of the decay of tetra-quark meson  $M_4 \to MM$,  Freund, Waltz and Rosner proposed a selection rule\cite{FWR}.   In SJM this FWR selection rule is interpreted as suppression  of junction-anti-junction hair pin diagram\cite{IO2} \cite{IOT}, similarly  to OZI rule\cite{O2}\cite{Z}\cite{Iiz} to suppress quark-anti-quark hair pin diagram. We will call the process shown in Fig. \ref{fig4}(b) as ``junction hair pin(FWR) rule" forbidden process. \par

\input epsf.tex


The junction hair pin(FWR) rule is viewed from a different standpoint. When we apply $1/N_c$ expansion\cite{tHooftN}\cite{Witten} to SJM, we obtain\cite{IOT_N} a suppression factor of  $O(1/N_c )$  for OZI rule and that of $O(1/\sqrt{N_c} )$ for junction hair pin(FWR) rule.  \par
We conclude this subsection by saying that (1)   $K^+n$ scattering  contains only $X$ type amplitude so that no resonance contribution is expected in this case due to duality and (2)  even if $\Theta$ exists,   $J\Jb$ hair pin line has to be newly created   in $K n\to\Theta$, which is a forbidden process. \par

\section { \bf Comments on parameters of SJM} 


\subsection { Mass of junction}\par

 One simple way to estimate junction mass $m_J$ is to use the mass of ordinary baryon and meson as follows,

\begin{eqnarray}
 m_B=3 m_q+m_J, 
\label{} 
\end{eqnarray}%
\begin{eqnarray}
m_M= 2 m_q.
\label{} 
\end{eqnarray}%
If we estimate $m_q$ from $\rho$ and $\o$ meson mass\footnote{Pseudoscalar meson should be discussed quite differently, since chiral symmetry and its spontaneous breaking plays essential role there.  } as $m_M= 2 m_q\sim ({3 m_{\rho}+m_{\o}})/4 =773$   \ MeV, $m_J\sim m_B-
3 m_q \sim 1172-1160 =12$   \ MeV, where $m_B$ is estimated from the average of nucleon and $\D_{33}$ resonance\footnote{$m_B\sim (4m_N+16m_\Delta)/{20}\sim 1172$   \ MeV.}.  The obtained value
\begin{eqnarray}
m_J\sim O(10) {\rm    \ MeV}
\label{massjunction} 
\end{eqnarray}%
is quite small, $m_J \ll m_B, m_M$.


\subsection{ Naive estimate of length of string from uncertainty relation.}\par
Every string in the ground state hadron in SJM is expected to have ``minimal length" due to zero point oscillation. The minimal length and the mass  are evaluated by uncertainty relation. 
Let $x$ be the distance from a junction to a source quark with mass $\tilde m$  and assume that the string connecting them is almost straight. The sum of  kinetic energy  of the source quark and linear potential energy to trap it is the constituent quark mass $m_q$, but we write it as $E$ in this subsection for the sake of convenience. \par
\begin{eqnarray}
%
E=\sqrt{p^2 +{\tilde m}^2} +\sigma \cdot x,
\label{En} 
\end{eqnarray}%
where $p$ is the source quark momentum and $\sigma$ is the string tension. \par
The string length  $L$ is estimated from $x$ such that it minimizes $E$ by taking $p= 1/x$ from the uncertainty relation. \par\noindent
(i) For light quarks ${\cal N}=(u, d)$ for which ${\tilde m} _{\cal N} \ll p$,  $E_{\cal N}$ is approximated as\par
\begin{eqnarray}
%
E_{\cal N} \sim p+(1/2){\tilde m}_{\cal N}^2/p+ \sigma\cdot  x .
\label{Enl} 
\end{eqnarray}%
%

\noindent 
By putting ${\tilde m}_{\cal N} =0$,  it   is minimized at $ x = 1/\sqrt{\sigma} $, so\par
\begin{eqnarray}
%
L_{\cal N}  \sim 1/\sqrt{\sigma} \quad  {\rm  and } \quad E_{\cal N} \sim 2\sqrt{\sigma}.
\label{Ln} 
\end{eqnarray}%
 When we set  the constituent quark mass $m_{\cal N} \equiv E_{\cal N} $ equal to $m_B/3$=0.390    \ GeV in the previous subsection by neglecting $m_J$, we obtain\par
\begin{eqnarray}
%
\sqrt{\sigma}\sim {\rm  0.195   \ GeV},
\label{} 
\end{eqnarray}%
 so that $L_{\cal N} \sim $ 5.1   \ GeV$^{-1}.$ \par\noindent
(ii) For heavy quarks $Q=(c, b, t)$ whose masses ${\tilde m}_Q \gg p$, $E_{Q} \sim {\tilde m}_Q + (1/2)(p^2/ {\tilde m}_Q) +\sigma\cdot  x $, which  is minimized at $x \sim 1/({\sigma} {\tilde m}_Q)^{1/3} $, so\par
\begin{eqnarray}
%
L_Q  \sim 1/({\sigma} {\tilde m}_Q)^{1/3} \quad   {\rm and} \quad  E_Q \sim {\tilde m}_Q \{1+(3/2)(\sigma/ {\tilde m}_Q^2)^{2/3}\}.
\label{} 
\end{eqnarray}%
Most of the constituent mass $m_Q \equiv E_Q$ comes from the source quark mass ${\tilde m}_Q$, and thus  the mass difference among constituent quark masses comes almost from that of the source quarks. \par

  It will be interesting to compare the above result with the one obtained by WKB approximation for $ m_Q \equiv E_Q$ trapped in one dimensional linear potential. 
  For linear potential\par
  \begin{eqnarray}
  V(x)=\left\{
             \begin{array}{ll}
                  
                  V_{0}      &; \qquad  |x|>a,\\ 
                   (V_{0}/a) |x|    &; \qquad  |x|<a,
             \end{array} \right. 
 \label{pot} 
\end{eqnarray}
\noindent where $V_{0}>0$.  WKB approximation gives\par
\begin{eqnarray}
%
E_{\rm WKB} \sim {\tilde m}_Q \{1 +( 9\pi^2 /128)^{1/3} (\sigma/{\tilde m}_Q^2 )^{2/3}\},
\label{WKB} 
\end{eqnarray}%
 \noindent  in the limit $V_{0} \rightarrow \infty, a \rightarrow \infty$, with  $ V_{0}/a =\sigma $  kept finite. Note that both $E_Q-{\tilde m}_Q$ and  $E_{\rm WKB}-{\tilde m}_Q$ have the same functional form with  different numerical factors:
$(E_Q-{\tilde m}_Q)\sim 1.7(E_{\rm WKB}-{\tilde m}_Q).$ 
\par
\noindent (iii) Intermediate of the above two cases is  strange quark, whose source quark mass is ${\tilde m_s}=O(1)\times \sqrt{\sigma} \sim p$.  
Although we have to resort to a numerical calculation finally, let us use the Taylor expansion of 
$E_s=p+(1/2){\tilde m}_{s}^2/p+ \sigma\cdot  x$ as Eq. (\ref{Enl}),  but  the second term ${\tilde m}_s^2/p$ is fixed in minimizing $E_s$ by  uncertainty relation.\footnote{For  ${\tilde m_s}\sim 0.13{\rm   \ GeV}\sim 0.7\times  \sqrt{\sigma}$,  we are able to confirm that ${\tilde m_s}/p$ with $p\sim \sqrt{\sigma}$ is  within convergence radius of Taylor expansion.}
 Then   $p\sim \sqrt{\sigma}$ and $E_s \sim 2{\sqrt \sigma}+ (1/2){\tilde m}_s^2/\sqrt{\sigma}.$\par
Here we get a somewhat surprising result as follows: 
The mass difference of  constituent quarks from the above is  $\Delta_s \equiv  E_s-E_{\cal N} \sim (1/2){\tilde m}_s^2/\sqrt{\sigma}$. The  numerical value $\Delta_s$=130$\sim$150    \ MeV is not obtained for ${\tilde m_s}\sim 130 \sim 150$    \ MeV. Namely, for   $\sqrt{\sigma}= 0.195$   \ GeV, we have $\Delta_s \equiv  E_s-E_{\cal N} \sim (1/2){\tilde m}_s^2/\sqrt{\sigma}\sim$  43    \ MeV for  ${\tilde m_s}=$130    \ MeV and $\Delta_s \equiv  E_s-E_{\cal N} \sim (1/2){\tilde m}_s^2/\sqrt{\sigma}\sim$  58    \ MeV for ${\tilde m_s}$=150    \ MeV. To reproduce the observed value of $\D_s\sim 130 \sim 150$ MeV, we have to take\par
\begin{eqnarray}
%
{\tilde m_s} \sim 2\Delta_s.
\label{ms} 
\end{eqnarray}%

Our numerical results are given in Table \ref{stringen}.  This is obtained from the minimum of $E$ of Eq. (\ref{En}) with $p=1/x$.  Table \ref{stringen} includes the case of Eq. (\ref{ms}).\par
%
\begin{table}
\caption[ ]{ String length and energy}
\label{stringen}
\begin{center}
\begin{tabular}{| c| c |c | c |c|} \hline 
quark  & ${\tilde m}_q$ (  \ GeV)& $L_q$(  \ GeV$^{-1}$)   &  $E_q$(  \ GeV)  & $ \D_q=E_q-E_{\cal N}$( GeV) \\ \hline
${\cal N}$  &0                &  5.13    &    0.390  &  0   \\ \hline
$s$      &  0.130            &  4.73    &    0.428  & 0.04    \\ \hline
${\bf s}$      &  {\bf 0.270}            &  4.18    &    0.520  &  {\bf 0.13}   \\ \hline
$c$      &  1.35             &  2.66    &    1.50   & 1.11   \\ \hline
\end {tabular}
\end{center}
\end{table}

For $s$ quark $L_s\sim 0.8 L_{\cal N}$. 
For heavy quarks $L_Q$ is much shorter. For example, $L_c \sim 0.5L_{\cal N}$ for charm quark, though the $1/x$ behavior  of the potential near $x \sim 0$ has to be taken into account of. \par

\subsection{ Comments on  parameter $\d$}\par

\begin{itemize}
\item[(1)]A direct way to estimate the parameter  $\d$ is to find an exotic meson $M^2_4$;
$$\d=2m_B-m(M^2_4).$$
We estimated it in 1977 as 
$ \d \sim 230 \sim 240{\rm    \ MeV}$
from the mass of $M^2_4$ candidate reported at that time\cite{IO1}\footnote{See Table IV (p.143) in section IV of Ref. \citen{IIMNOST}.}.  See, however, the next subsection newly added. \par

\item[(2)]${\Bb}B$ duality\par
An indirect way to estimate  $\d$ is to consider the  ${\Bb}B$ duality. Above the threshold of ${\Bb}B$-channel, the  large  difference $\D \s_T$ of Eq.(\ref{delsigB}) may imply  that numbers of $M^2_4$ above the threshold contribute to ${\rm Im} T({\pb} p)$,  so that  $\d \sim$ small positive or negative. \par
\end{itemize}

\subsection{Tetra-quark meson of Belle experiment,  parameter $m_{IJ}$ and $\d$ in SJM}\par

The Belle Collaboration  has reported \cite{Belle} a new resonance $Z^+(4430)$ in the invariant mass distribution of $\pi^{\pm} \psi'$. We consider this  $Z^+(4430)$  is a good candidate for $M^2_4(uc J ;\Jb \cb\db)$ in SJM. By this assignment 
we can ``directly" determine the parameter $m_{IJ}$, energy of inter-junction string, and thus $\d$. \par
The constituent quark mass $m_q$ is set equal to $E_q$ in Table \ref{stringen}   as: 
\begin{eqnarray}
%
m_{\cal N}\equiv E_{\cal N}, \quad m_s\equiv E_s 
 \quad{\rm and}\quad m_c\equiv E_c. 
\label{} 
\end{eqnarray}%
%
If we adopt 4430   \ MeV as an input for the mass of  $M^2_4(uc J ;\Jb \cb\db)$, we can set
\begin{eqnarray}
%
2m_{\cal N}+2m_c+m_{IJ}=4430\ {\rm MeV (input}).
\label{input} 
\end{eqnarray}%
We obtain
\begin{eqnarray}
%
m_{IJ}=4430-780-3000=650{\rm    \ MeV}.
\label{mIJ} 
\end{eqnarray}%
In the estimate, junction mass $m_J$  and source quark masses of $u, d$ quarks are neglected. We also find 
\begin{eqnarray}
%
\d= 2m_{\cal N}-m_{IJ}=780-650=130{\rm    \ MeV}.
\label{} 
\end{eqnarray}%
By these  $m_{IJ}$ and $\d$ and  values of Table \ref{stringen}, we  give  masses of unconventional hadrons. Especially, mass of penta-quark baryon $\Theta=B^3_5(ud J;\sb \Jb; J ud)$ is
\begin{eqnarray}
%
m(\Theta(B^3_5(ud J;\sb \Jb; J ud))) =4m_{\cal N}+m_s+2m_{IJ}=3380{\rm    \ MeV},
\label{} 
\end{eqnarray}%
which is much heavier than the value given by Skyrmion model\cite{Diak}\cite{Prasz} or di-quark model\cite{JW}.
Masses of various unconventional hadrons are listed below in MeV.\par
\noindent{\bf Unconventional meson} $M^2_4$:\par
%
\begin{eqnarray}
%
\begin{array}{ll}
 Z^+=M^2_4(uc J; \Jb\cb \db)  &=2m_{\cal N}+2m_c+m_{IJ}=4430({\rm input})\\
M^2_4(\Nc\Nc J ;\Jb\Ncb \Ncb)  &=4m_{\cal N}+m_{IJ}=2210,  \\ 
  M^2_4(u s J; \Jb  \sb \db)  &=2m_{\cal N}+2m_s+m_{IJ}=2470  .
\end{array}
\label{} 
\end{eqnarray}%
{\bf Unconventional meson} $M^2_2$:\par
%
%
\begin{eqnarray}
%
\begin{array}{ll}
 M^2_2(\Nc J ;\Jb \Ncb)  &=2m_{\cal N}+2m_{IJ}=2080\\ 
 M^2_2(c J ;\Jb \cb)  &=2m_{c}+2m_{IJ}=4300. 
\end{array}
\label{M22} 
\end{eqnarray}%
{\bf Unconventional meson $S^2_0$(flavor singlet gluonium)}:
\begin{eqnarray}
%
S^2_0( J ;\Jb)=3m_{IJ}=1950.
\label{} 
\end{eqnarray}%
{\bf Unconventional baryon} $B^3_5$:
%
\begin{eqnarray}
%
\begin{array}{ll}
 B^3_5(\Nc\Nc J; \Ncb \Jb; J \Nc\Nc)  &=5m_{\cal N}+2m_{IJ}=3250,  \\ 
 \Theta=B^3_5(ud J;\sb \Jb; J ud)  &=4m_{\cal N}+m_s+2m_{IJ}=3380.
\end{array}
\label{} 
\end{eqnarray}%
{\bf Unconventional baryon} $B^3_3$:

%
\begin{eqnarray}
%
\begin{array}{ll}
 B^3_3(\Nc\Nc J;\Jb J \Nc)  &=3m_{\cal N}+3m_{IJ}=3120.
 \end{array}
\label{} 
\end{eqnarray}%
{\bf Di-baryon} $D^4_6$:
\begin{eqnarray}
%
D^4_6(\Nc\Nc J ;\Nc\Nc J ;\Jb ;\Nc\Nc J)=6m_{\cal N}+3m_{IJ}=4290.
\label{} 
\end{eqnarray}%

In the Belle Collaboration, decay width of $Z^+(4430)$ is reported as ``relatively narrow":
\begin{eqnarray}
%
\Gamma=45{\rm    \ MeV}.
\label{gamma} 
\end{eqnarray}%
In SJM,  decay $Z^+(4430)\to \pi \psi'$ occurs only through  forbidden process by  junction hair pin(FWR) rule(see Fig.\ref{fig4}(b)). \par

\par

\section { \bf Summary } 


 We consider the junction as not merely a mathematical symbol but a physical entity playing a role of connector of orientable strings, and discussed the nature of unconventional hadrons    with  skeleton-like structure of junctions. We  introduce cluster hypothesis into SJM. Based on this picture, mass of hadrons with multiclusters is given,
%
\begin{eqnarray}
%
\begin{array}{rl}
  m =&m_q N_q+m_J N_J+m_{IJ} N_{IJ}\cr  
    =&m_B \cdot N_J-\delta\cdot N_{IJ}.
\end{array}
\label{} 
\end{eqnarray}%
Parameters $m_q $, $m_J$, $ m_{IJ}$ are estimated:
\begin{itemize}
\item[1)]Minimal length  and energy $m_q $ of string with quark at the end are estimated by uncertainty relation in section 4.2.\par
\item[2)]
We estimate the mass of junction   as
\begin{eqnarray}
m_J\sim  O(10)  {\rm     \ MeV},
\label{mj} 
\end{eqnarray}%
which is quite small: $m_J \ll m_B, m_M$. 
Among many papers based on AdS/CFT which cite some sketches of string junction structure of ordinary baryon, we are interested in Imamura's paper that evaluates $m_{J}$  to be quite small\cite{Imamura}\cite{GI}.

\par
\item[3)]
By identifying  $Z^+(4430)$ with   $M^2_4(uc J; \Jb \cb\db)$, we  estimate the inter-junction energy   $m_{IJ}$ and the  energy $\d$ to cut the inter-junction string in subsection 4.4,
\begin{eqnarray}
%
m_{IJ}=650 \ {\rm MeV}  \quad {\rm and}\quad \d=130 \ {\rm MeV}.
\label{mIJd} 
\end{eqnarray}%
Investigation of inter-junction energy, $m_{IJ}$, based on AdS/CFT and lattice gauge theory will be awaited.\par
\item[4)]Taking into account the smallness of  $\d$ and $m_J$, the mass of hadrons in SJM is given approximately by $N_J\cdot m_B $.   Together with  parameters given in Table \ref{stringen}, 
  masses of  hadrons including unconventional ones are predicted as in 4.4. For example,  $ \Theta=B^3_5(ud J;\sb \Jb; J ud)  =4m_{\cal N}+m_s+2m_{IJ}=3380$ MeV.  This value is larger than those  given by Skyrmion model and di-quark model. \par
\end{itemize}
\vskip 0.5cm 
 
Recently one of the authors(S. O.) was informed from Tamagaki\cite{Tama} that he is making study 
about such universal  repulsion among three  baryons that is necessary to 
stabilize neutron stars under the mixture of hyperons. This is the extension of his work about the universal repulsion between two  baryons(R. Tamagaki, Bulletin of the Institute of for Chemical Research, Kyoto  Univ. , 
Vol. 60, No.2(1982),190).
By noting flavor independence of these two universal repulsions, he attributes them to  flavor independent nature of the junction  of SJM.\par

\section*{Note Added}
In the previous version of the paper, there was a mistake in the description related to penta-quark baryon belonging to $\overline{10}_f$. We  corrected the presentation related to this point.\par

\section*{Acknowledgements}
Preliminary version of this paper was presented by one of the authors(M. I.) at the annual meeting in Department of Physics, Yamagata University on April 22, 2004.
We are stimulated by the work of Tamagaki. We would like to express our deep acknowledgement to him.  We are grateful to Ishihara and Ghoroku for discussions on AdS/CFT approach.\par


\end{document}